\documentclass[%
 reprint,
 amsmath,amssymb,
 aps,
]{revtex4-2}

\usepackage{graphicx}
\usepackage{dcolumn}
\usepackage{bm}
\usepackage{cleveref}
\usepackage{subcaption}
\begin{document}

\preprint{APS/123-QED}

\title{Final velocity and radiated energy in numerical simulations of binary black holes}

\author{Emmanuel A. Tassone}
  \email{emmanuel.tassone@unc.edu.ar}
\author{Carlos N. Kozameh}
 \email{carlos.kozameh@unc.edu.ar}
\affiliation{
 FaMAF, Universidad Nacional de C\'ordoba\\
\normalsize \em \small 5000, C\'ordoba, Argentina 
}%

\date{\today}

\begin{abstract}
The evolution of global binary black holes variables such as energy or linear momentum are mainly obtained by applying numerical methods near coalescence,post-Newtonian expansions or a combination of both. In this paper, we use a fully relativistic formalism presented several years ago that only uses global variables defined at null infinity together with the gravitational radiation emitted by the source to obtain the time evolution of such variables for binary black holes (BBH) systems. For that, we use the Rochester catalog composed of 776 BBHs simulations. We compute the final velocity, radiated energy, and intrinsic angular momentum predicted by the dynamical equations in this formalism for non-spinning, aligned/anti-aligned spins, and several different precessing configurations. We compare obtained values with reported values in numerical simulations. As BBHs parameter space is still not completely covered by numerical simulations, we fit phenomenological formulas for practical applications to the radiated energy and final velocities obtained. Also, we compare the fits with reported values. In conclusion, we see that our formulae and correlations for the variables described in this work are consistent with those found in the general literature.
\end{abstract}

\maketitle

\section{\label{Introduction}Introduction}
The RIT Catalog of Numerical Simulations \cite{healy2020} is a catalog containing the gravitational wave data for 776 numerical relativity simulations. The catalog spans a large variety of initial conditions and provides the gravitational strain for the evolution of the mergers during all the phases.

It is believed that spin orientation on the initial state of the binary black hole (BBH) can constrain the evolutionary processes that lead the binary to the merger \cite{hemberger2013}. Spin orientations can influence the kick velocities \cite{gonzalez2007}, making them high enough for a merged binary to be ejected from the nucleus of a massive host galaxy \cite{campanelli2007,campanelli2007maximum}. The probability of these large recoils depends on the distribution of mass ratios and spins of the progenitor binaries \cite{healy2014}. Spin orientations also influence the energy radiated, and particularly, the maximum luminosity of the merger\cite{hemberger2013}. Additionally, there is also research made on the \textit{hangup effect}, which is a change in the merger duration compared to non-spinning binaries due to the presence of spins, resulting in faster or slower fusion of the binaries \cite{healy2018hangup,campanelli2006spinning}.

Other important parameters when setting up a black hole simulation are the masses of the black holes which are represented typically by the quotient  $q=m_1/m_2$, with $m_1$ being the smaller black hole progenitor. The mass quotient, along with the spin orientation, may determine the remnant mass, spin and recoil velocity \cite{zlochower2015}.

Although the spectrum of parameters along all the 776 simulations in Rochester catalog is wide, it is still computationally prohibitive to cover the whole BBH parameter space exhaustively, making fitting formulas useful for practical purposes. Many fitting formulas have been deduced that relate the final spin of the remnant black hole with the initial spins and mass parameters \cite{rezzolla2008final,hofmann2016final,lousto2014black,lousto2010remnant}. However, most of the relations established from the data analysis are constructed based on models from the parametrized post-Newtonian formalism (PPN). The PPN formalism allows finding approximate solutions to the Einstein field equations for the metric tensor, and therefore, obtaining physical quantities such as energy, velocity and angular momentum when the binaries are far away from each other\cite{blanchet2014gravitational}.

In this paper, we perform the numerical evolution of four global physical variables defined at null infinity, namely, the Bondi energy and linear momentum, the center of mass velocity and intrinsic angular momentum. The last two variables are defined using a formalism developed by Kozameh and Quiroga (KQ)\cite{kozameh2016}. 

The evolution of the before-mentioned variables is made using the Rochester gravitational wave strain data catalog. We analyze the dynamics of the final state of the resulting black hole and relate it to the most significant variables before the coalescence. Hence, the initial spin orientations, orbital angular momentum, initial masses, maximum energy lost, final total velocity, final total angular momentum and its tilting angle are going to be studied in this work. Many interesting correlations between these variables are shown in the work.

This paper is organized as follows. In Sec. \ref{Background}, we describe the equations of motion used for this work. In Sec. \ref{Numerical framework}, we mention some aspect of the code employed to compute the evolution and the subsequent data analysis. In Sec \ref{Analysis by groups}, we analyze the results obtained for the final states of the evolution. Then we make a deeper analysis, classifying the results into three categories according to the kinematics of the BBH spins: non spinning (NS), aligned (A) and precessing (P). There are 29 simulations in first category (NS), 447 with aligned spins (A) and 300 simulations are precessing spins BBH (P). Finally, Sec. \ref{Discussion} closes the work with a discussion and conclusion about the most relevant aspects found and also a discussion over the formalism employed and potential future work.

\section{\label{Background}Background}
The KQ formalism \cite{kozameh2016} provides a fully relativistic treatment that defines the center of mass frame and intrinsic angular momentum using the asymptotic symmetries represented by the Winicour linkages \cite{held1980:3}. It is worth mentioning that a different formulation based on null congruences with twist by Adamo, Newman and Kozameh (ANK) \cite{adamo2012null} yields similar evolution equations at a quadrupole level. In this sense, the KG formulation has been shown to be consistent with PN results up to octupole terms in the gravitational radiation \cite{kozameh2018}.

The physical quantities for asymptotically flat spacetimes are given by the Weyl scalars $\psi_0$, $\psi_1$, $\psi_2$, $\psi_3$, and $\psi_4$ at null infinity. Those scalars are defined on special reference frames called Newman-Unti coordinates \cite{newman1963class}. Each Newman-Unti foliation is then associated to an arbitrary timelike wordline on a fiducial flat space, called the Observation space, where the equations of motion are given and numerically solved. These special frames allow us to define physical variables on "non-inertial" frames. Using the Winicour linkages together with a gauge fixing procedure, the KG formalism defines the center of mass and intrinsic angular momentum. Using the Bianchi identities at null infinity, one obtains the evolution equations for those variables on the Observation space. The reader may refer to \cite{kozameh2016,kozameh2018,kozameh2020} for a thorough construction of the physical quantities using the KG formalism.
The relevant evolution equations for the purpose of this work read
\begin{align}
	\dot{M}&=-\frac{c}{10\sqrt{2}G}(\dot{\sigma}_{R}^{ij}\dot{\sigma}_{R}^{ij}+\dot{\sigma}_{I}^{ij}\dot{\sigma}_{I}^{ij})\label{Mdot},\\
	\dot{P}^{i}&=\frac{2c^{2}}{15\sqrt{2}G}\dot{\sigma}_{R}^{jl}\dot{\sigma}_{I}^{kl}\epsilon^{ijk}\label{Pdot},\\
\dot{D}^{i}&= \sqrt{2} P^{i}+ \frac{3c^2}{7G}[\dot{\sigma}_{R}^{ijk}\sigma_{R}^{jk}+\dot{\sigma}_{I}^{ijk}\sigma_{I}^{jk}- \sigma_{R}^{ijk}\dot{\sigma}_{R}^{jk}- \sigma_{I}^{ijk}\dot{\sigma}_{I}^{jk}],\label{Ddot}\\
\dot{J}^{i}&=-\frac{c^{3}}{5G}(\sigma _{R}^{kl}\dot{\sigma}_{R}^{jl}+\sigma _{I}^{kl}\dot{\sigma}_{I}^{jl})\epsilon^{ijk}, \label{Jdot}
\end{align}
where $M^{i}$ and $P^{i}$ are the total Bondi mass  and linear momentum respectively, $D^{i}$ is the dipole mass momentum, and $J^{i}$ the total angular momentum. The first two equations yield the evolution of the Bondi four-momentum whereas the last two yield the corresponding evolution of the dipole mass momentum and total angular momentum. These last two variables are the components of the so-called relativistic angular momentum  2-form and behave like the electric and magnetic fields under a Lorentz transformation. The symbols  $\sigma_R$ and $\sigma_I$ denote the real and imaginary part of the Bondi shear at null infinity. Commonly, the Bondi shear is denoted as $\sigma_0$ and accounts for the deformation of the spacetime due to the gravitational radiation that arrives at null infinity. Usually $\sigma^0$ is expanded in terms of spin-weighted spherical harmonics $Y^s_{lm}$, very useful for analyzing gravitational waves \cite{thorne1980multipole}. Yet, the KQ formalism expands the asymptotic shear $\sigma^0$ in terms of tensor spherical harmonics $Y^s_{i_1 i_2...}$ \cite{newman2005}. This base is fundamental to define covariant quantities as in \cref{Mdot,Pdot,Ddot,Jdot}. The dot on the $\sigma^{ij}$ stands for the derivative with respect to the Bondi time coordinate at  future null infinity. It is worth mentioning that in the above equations we have omitted quadratic octupole terms and all remaining coefficients for $l\geq 4$ since they are negligible in BBH coalescence.

Finally we write down the equations for the center of mass $R^{i}$ and intrinsic angular momentum $S^{i}$ as given in the KG formalism. They are related to the previous variables by 
\begin{align}
D^{i}&= M R^{i}+c^{-2} \epsilon^{i j k} \frac{P^{j}}{M} S^{k}-\frac{8}{5 \sqrt{2} c} P^{j} \Delta \sigma_{R}^{i j}\label{D}\\
J^{i}&=S^{i}+\epsilon^{i j k} R^{j} P^{k}-\frac{137c^2}{168 \sqrt{2}G}(\sigma_{R}^{ijk}\sigma_{I}^{jk}- \sigma_{I}^{ijk}\sigma_{R}^{jk})\label{J}
\end{align}

The center of mass $R^i$, is the spatial component of a covariant vector of the formalism and represents the physical analogue of the newtonian center of mass but also accounting for the radiated energy. One can take the derivative with respect to the Bondi time $u$ and define the center of mass velocity $V^i$ as the change of the center of mass with respect to the Bondi time. One of the purposes of this work will be to show that the center-of-mass velocities given by the KQ formalism agree very well with the recoil velocities found in the RIT catalog.

The equation for $J^i$ resembles its newtonian counterpart except for the extra radiative terms. In this work the last term is negligible for the BBH catalog but it could be important in other situations where the octupole term is not negligible. A fully relativistic version of these equations can be found in reference \cite{kozameh2020}. Note that the orbital angular momentum can be important when the center of mass is displaced from the origin and achieves high velocities due to the gravitational kick.

\section{\label{Numerical framework} Numerical framework}
The numerical code employed to process data and run the evolution equations is described in this section. The code can be accessed at \cite{code}.

The Rochester catalog has been used before to obtain the dynamical evolution of several physical observables using the KG formalism \cite{tassone2021}. However, the latter work has a numerical error associated to the strain in the last steps of the evolution. This error is due to a well-known effect of performing a raw numerical integration \cite{reisswig2011notes}. Although the numerical error was small enough to keep the qualitatively analysis of the physics still valid, in this work we corrected the code so as to avoid the linear drifts coming from the integration on sigma.

The amplitude and phase for the simulations are obtained in compressed format from the Rochester catalog \cite{RITcatalog}. An interpolation is made with one dimensional smoothing splines. Order k=5 splines are implemented in the interpolation. Subsequently, the polarizations of gravitational $h^{lm}_{+}, h^{lm}_{\times}$ wave strain can be obtained by the relation,
\begin{equation*}
    r h^{lm}_{+}-i r h^{lm}_{\times}=A_{lm}e^{i\phi_{lm}},
\end{equation*}
where $r$ is a radial coordinate, $l$ and $m$ the modes of the gravitational wave, $A_{lm}$ and $e^{i\phi_{lm}}$ the amplitude and phase of the gravitational wave respectively. Asymptotically, the strain and shear are directly related by the $\Psi_4$ scalar, implying
\begin{align*}
    h_{+}=-\sigma_R,\\
    h_{\times}=-\sigma_I.
\end{align*}
The code implements the calculation of the shear and then it transforms from the spinor to the tensor harmonics basis using the transformation relations in \cite{mandrilli2020}. After that, the calculation for all the physical variables is implemented. In particular, the evolution of \cref{Jdot,Mdot,Pdot} can be carried out for all times. Finally, the code implements functions to collect and analyze the results obtained. Data analysis is shown in Sec. \ref{Analysis by groups}.

\section{\label{Analysis by groups} Main results}
In this section we first describe the general results obtained for the global variables of an asymptotically flat space time containing a BH binary system. We provide the distribution of the data and give an overview of the physics obtained for all the simulations. Then, we dedicate a subsection for each categories made of the Rochester repository: non-spinning Binaries, aligned-spins binaries and precessing binaries.

We made a special analysis of the variables obtained by \cref{Mdot,Pdot,Ddot,Jdot,D,J} and the correlations with initial parameters from the simulations, such as initial total angular momentum $J_{in}$ and mass ratio $q$. It is important to distinguish these two sets of parameters. The first set, the global variables, are well-defined in our formalism and can be obtained at null infinity without knowledge of the BH masses, spins and orbital angular momentum. For instance, to obtain $J_{in}$ we only need the knowledge of the center of mass position and the definition of intrinsic angular momentum.

The second set of parameters corresponds to all the initial conditions contained in the Rochester repository, locally defined, containing the initial spins, masses, and orbital angular momentum of the BH binaries. Knowledge of this second set is very important to analyse the behaviour of BH coalescence, and gives further understanding of this kind of isolated system.

Throughout the following subsections we will distinguish between six classes of binaries: equal mass non-spinning (EM-NS), equal mass aligned (EM-A), equal mass precessing (EM-P), non-equal mass non-spinning (NEM-NS), non-equal mass aligned (NEM-A), and non-equal mass precessing (NEM-P).
\subsection{General results}\label{General results}
\begin{figure}[htb]
	\centering
	\includegraphics[scale=0.5]{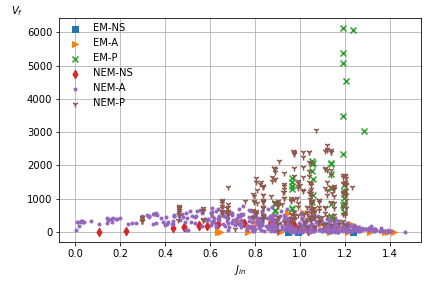}
	\caption{Correlation between the final center of mass velocity, $V_f$ (in $km/s$), and the initial total angular momentum, $J_{in}$.}
	\label{fig1}
\end{figure}

The results for the final center of mass final velocity $V_f$ and initial total angular momentum $J_{in}$ are presented in Fig. \ref{fig1}. The final velocity $V_f$ is defined as,
\begin{equation}\label{V_final}
    V^i_f=\dot{R}^i_f,
\end{equation}
where $\dot{R}^i_f=\dot{R}^i(t_f)$ is the derivative with respect to Bondi time evaluated at the final time of the simulation and the index $i$ denotes the component of the vector. 

The plot in Fig. \ref{fig1} shows the final velocities for the 776 simulations in the catalog. The vertical axis $V_f$ is the norm of the vector velocity defined in Eq. (\ref{V_final}). One can see that precessing binaries exhibit higher final velocities. This observation could be relevant both at a level of predicting the final velocities of the residual black hole or at the level of observational data. If the final velocity of the residual black hole can be observed (via Doppler effect of a surrounding coalescing gas), then receding speeds higher than $\sim 750\ km/s$ implies that the original binaries had precessing spins. One can even speculate that if the final speeds are higher than $\sim 3000\ km/s$, then the initial binaries were in the class of EM-P. It is also possible to recognize in Fig. \ref{fig1}, that within each class, there is a value for the initial total angular momentum $J_{in}$ with the highest final velocity $V_f$. The distribution of the simulations seems to exhibit an asymmetric peaked-like distribution, whose peak value varies among each subclass in the classification. On account of its symmetry, the EM-NS class has vanishing final velocities (see formula (\ref{Fitchett}) below). The reader should be aware these assertions are rather approximate due to the bias in the RIT catalog parameter space. However, the parameter space is vast enough to visualize the claims made in the paragraph and in this sense phenomenological formulae are fitted in the following sections with the idea of filling the blanks in the initial parameter space. A special exception must be made with precessing binaries, for which the final state is known to be chaotic with respect to the initial parameters.

\begin{figure*}[htb]
	\centering
	\begin{subfigure}{.32\textwidth}
	\centering
	\includegraphics[width=\linewidth]{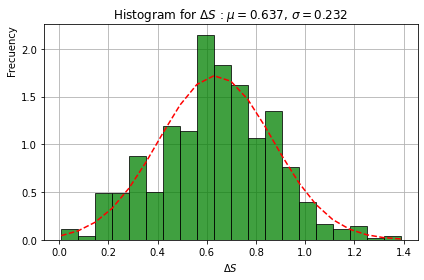}
	\caption{Distribution of angular momentum change $\Delta S$ (in code units). }
	\label{fig2:a}
	\end{subfigure}
    \begin{subfigure}{.32\textwidth}
    \centering
	\includegraphics[width=\linewidth]{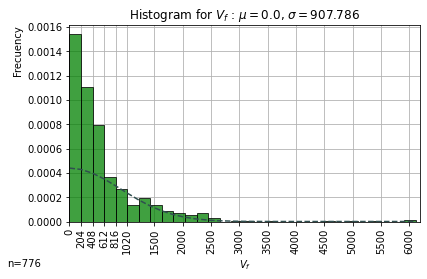}
	\caption{Distribution of final velocity $V_f$ (in km/s)}
	\label{fig2:b}
	\end{subfigure}
	\begin{subfigure}{.32\textwidth}
	\centering
	\includegraphics[width=\linewidth]{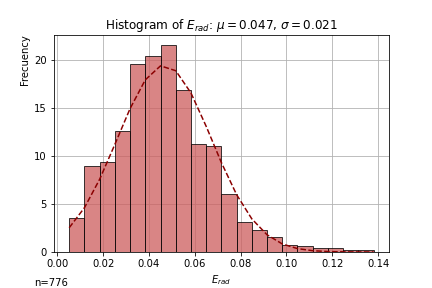}
	\caption{Distribution of total radiated energy $E_{rad}$}
	\label{fig2:c}
	\end{subfigure}
	
    \caption{Bar plots of the distribution from main variables of interest obtained after evolution in all the simulations.}
	\label{fig2}
\end{figure*}

We have solved the \cref{Mdot,Ddot,Pdot,Jdot,D,J} and plotted the data distribution for the variables $E_{rad}$,$V_f$ and $\Delta S$ in form of histogram (Fig.\ref{fig2}). Again, the variables of distributions in Fig.\ref{fig2} are also biased by the selection of simulations in RIT catalog. All the distributions are normalized as a probability density function and a gaussian function is fitted to get representative values. For the final velocity in Fig.\ref{fig2:b}, note that most of the simulations will end with kicks less than $\sim 1000\ km/s$. It is our purpose to investigate in the following sections which are the factors that cause this large variation of final velocities. Something important to remark is that final velocities are expressed in Bondi time used in our formalism. For comparison purposes, we should divide our final velocities by a $\sqrt{2}$ factor. This factor accounts for the transformation from the Bondi time $u$ to the standard time $t$ due to have defined $u=\frac{t-r}{\sqrt{2}}$ and not $u=t-r$ as found in common literature. Consequently, the kick velocities in terms of the coordinate time $t$ are lower than those showed in Fig. \ref{fig2:b}. 

It is more convenient to plot the data distribution for a normalized radiated energy instead of the raw energy change, as in Fig. \ref{fig2:c}. We define the total energy radiated as
\begin{equation}\label{Erad}
    E_{rad}=1-\frac{M_f}{M_i},
\end{equation}
where $M_i$ is the initial ADM mass of the BBH and $M_f$ the final Bondi mass of the remnant obtained by Eq. (\ref{Mdot}). Note the difference between the final $M_f$ in this work and the commonly used Christodoulou mass in the numerical literature. The distribution of radiated energy for our set of solutions is found between the range $\sim 0-13\%$. The histogram is normalized and the mean of the radiated energy is $\mu=0.047$ and the dispersion $\sigma=0.021$.

\begin{figure}[htb]
	\centering
	\includegraphics[scale=0.5]{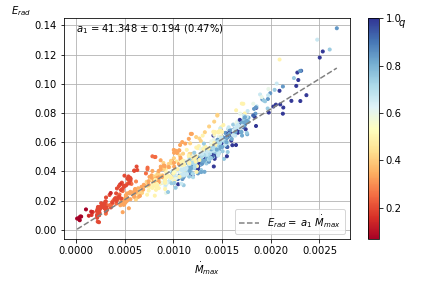}
	\caption{Correlation between total radiated energy $E_{rad}$ and the peak radiated energy $\dot{M}_{max}$ for all simulations. Dashed line shows fit to data. Colorbar indicates the value of mass ratio $q=m_1/m_2$.}
	\label{fig3}
\end{figure}

Definition (\ref{Erad}) is directly proportional to the peak energy loss $\dot{M}_{max}$ for binary coalescence. The bigger the peak energy loss, the bigger will also be the total energy radiated. In Fig. \ref{fig3}, we fitted a linear function of the form $E_{rad}=a_1 \dot{M}_{max}$ to the relation and found a proportionality constant 
\begin{equation}
    a_1=41.348 \pm 0.194
\end{equation}
with an error of $0.47\%$. Moreover, the proportionality relation could be made more precisely if fitted for each mass ratio as seen in Fig. \ref{fig3}. Having this result in mind, figures present in this work with $E_{rad}$ variable will be nearly the same as figures with $\dot{M}_{max}$ variable but with a change of scale. It must be emphasized that this proportionality relation is a feature of the BBH coalescence and will not be valid in general. We analyze these variables for each subclass in the following sections more specifically.

\begin{figure}[htb]
	\centering
	\includegraphics[scale=0.5]{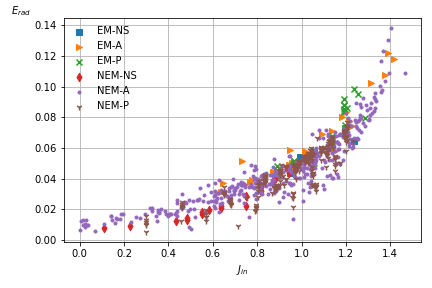}
	\caption{Correlation between the radiated energy $E_{rad}$ and the initial angular momentum $J_{in}$ for all the simulations}
	\label{fig4}
\end{figure}
To end this section, we also give an overview of the distribution of radiated energy with respect to the initial total angular momentum $J_{in}$ in Fig.\ref{fig4}. $E_{rad}$ vs $J_{in}$ clearly shows a non-linear relationship. This feature will be analysed for each group made in the next subsection and also in the discussion. A model to understand the behaviour of the radiated energy with respect to the initial total angular momentum is developed in Appendix \ref{AppendixA}. It is worth mentioning that this model has a post-Newtonian derivation. Thus, we will contrast the fully relativistic numerical evolution of variables that have asymptotic nature with a PN model describing its dynamics. It is rather remarkable that with few phenomenological parameters that are adjusted to fit the numerical data, the PN formula matches the numerical data.

\subsection{Non-spinning binaries}\label{Non-spinning binaries}
In this subsection we describe the group of simulations which have no initial spins. This group of binaries has the simplest physics to describe since the space of initial parameters has the lowest dimension. 

As a first step, we calculate the correlations of the final velocity and the maximum gravitational luminosity of the binary system with the mass ratio $q$. The gravitational radiation is calculated by Eq.(\ref{Mdot}) and it is directly related to the luminosity by integrating it into a spherical surface area far away from the source. As the final luminosity is a quantity that can be usually measured, it is relevant to know if a direct relation exists between the radiation per unit time with the mass ratio of the BBH. Likewise, assuming the final velocity of the remnant lack hole can be observed, the correlation with $q$ also yields information about this particular coalescence.

\begin{figure}[htb]
	\centering
	\includegraphics[scale=0.5]{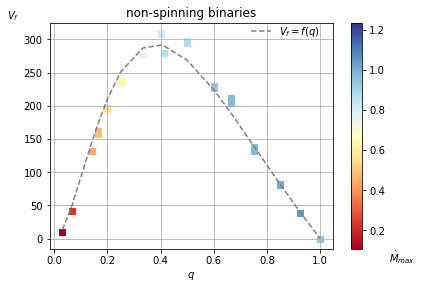}
	\caption{Correlation between the final velocity $V_f$ and mass ratio $q$. Colorbar indicates the maximum energy lost $\dot{M}_{max}$. Dashed line shows the Fitchett model fit.}
	\label{fig5}
\end{figure}

Correlations between $V_f$, $q$ and $\dot{M}_{max}$  are outlined in Fig. \ref{fig5}.  We use the Fitchett’s recoil model for circular orbits proposed in \cite{fitchett1983influence} whose dependence with q is
\begin{equation} \label{Fitchett}
    V_f(q)= a\ \frac{q^2 (1-q)}{(1+q)^5}.
    \end{equation}
with $a$ a fitting constant. We compute the parameter $a$ by a least square method. The best fit to our data has the coefficient $a= 16317.78 \pm 207.93\ km/s $. The derivation of the Eq. (\ref{Fitchett}) is also given in Appendix A.

We found the maximum final velocity $V_f=309.43$ reached at the value $q=0.4$. Null velocities are reached for limit cases when $q=0$ or $q=1$, i.e. mass $m_1=0$ or $m_1=m_2$. This is consistent with the fact that when $q=0$, the center of mass is in the center of the mass $m_2$; hence, it does not move while the mass $m_1$ goes to zero. Similarly, when $q=1$, the center of mass does not move either due to the symmetry of the problem and coincides with half the distance between the binaries. 

Even though limit cases of $q$ yield vanishing velocities, Fig. \ref{fig5} shows that while $q=0$ has no energy loss, $q=1$ allows the most radiating systems. Thus,  $q=1$ non-spinning binaries systems are most likely to be detectable by its higher luminosity, whereas intermediate values for $q$ are most likely to be detectable by their kick velocities. 

\begin{figure}[htb]
	\centering
	\includegraphics[scale=0.5]{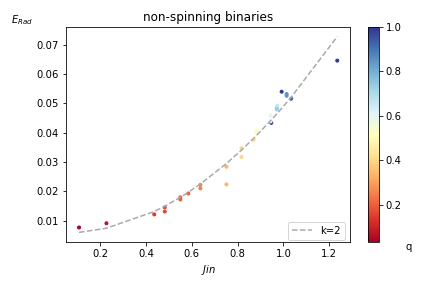}
	\caption{Correlation between the total radiation emitted $E_{rad}$ and the initial total angular momentum $J_{in}$. Colorbar indicates different levels of the initial mass ratio $q$. Letter $k$ indicates polynomial degree of the fit.}
	\label{fig6}
\end{figure}

Another interesting relation we have explored is the radiated energy $E_{rad}$ vs the initial total angular momentum $J_{in}$. For the NS subgroup, this plot yields Fig \ref{fig6}.

Fig.\ref{fig6} is fitted with a quadratic curve of the form
\begin{equation} \label{Erad=bJin^2}
    E_{rad} = b\ J_{in}^2,
\end{equation}
with $b=0.050 \pm\ 0.001$. The model fitted in  Eq. (\ref{Erad=bJin^2}) is motivated by the derivation made in Appendix \ref{AppendixA}.

Likewise, we can also give the correlation between $\dot{M}_{max}$  and $q$. The plot is given in Fig. \ref{fig6.5} and the phenomenological formula containing two parameters is derived in Appendix \ref{AppendixA}. The relation is given as,
\begin{equation}\label{Lum-max}
	\dot{M}=-\frac{A}{1000}\frac{q^2}{(1+q)^4}\left( 1 + \frac{B}{216}\left(\frac{1-q}{1+q}\right)^2\right),
\end{equation}
 with $A$ and $B$ to be obtained from the numerical data. Notice that the formula has a PN background as can be checked in Appendix \ref{AppendixA}. It follows from the data or the formula that the luminosity has a minimum at $q=0$ and a maximum at $q=1$, i.e., equal BBH masses yield maximum luminosity. The coefficients found with least squares method are
 
 \begin{align}
    A&=22.44 \pm 0.06\ (0.30\%) \label{coeffA}\ ,\\
    B&=-159.00 \pm 4.74\  (2.98\%)\label{coeffB}\ . 
 \end{align}
 
 Note also that relationship (\ref{Lum-max}) is injective. Thus, given the coefficients in \cref{coeffA,coeffB}, one can invert the formula (\ref{Lum-max}) and determine which mass ratio $q$ corresponds to a particular value of the measured luminosity, i.e., global dynamical information is used to obtain local information of the BBH.
 
\begin{figure}[htb]
	\centering
	\includegraphics[scale=0.5]{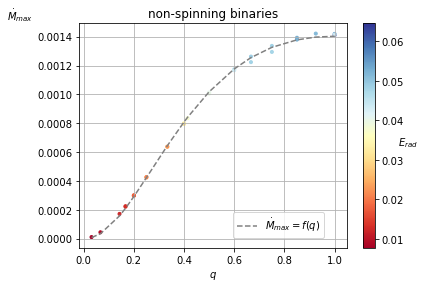}
	\caption{Correlation between the peak energy radiated $\dot{M}_{max}$ and mass quotient $q$. Colorbar indicates values of energy radiated which are proportional to the peak energy loss. Dashed line shows the model fit to data.}
	\label{fig6.5}
\end{figure}

\subsection{Aligned-spins binaries}\label{Aligned binaries}

In this section, we analyse the aligned-spin class. The initial spins in this kind of configurations are likely to remain in the z-direction as aligned-spins binaries are known to be stable configurations \cite{kozameh2020spin}. These kinds of binaries are more realistic ones and hence relevant for astrophysical applications.

\begin{figure*}
	\centering

	\begin{subfigure}{.49\textwidth}
		\centering
		\includegraphics[width=\linewidth]{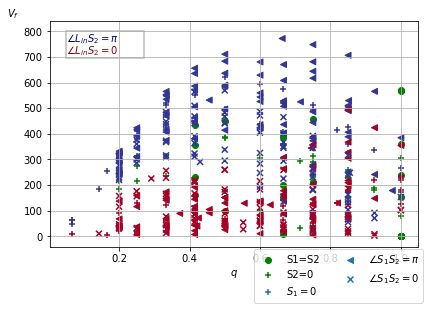}
        \caption{Red and blue colors denote alignment of spin $S_2$ with respect to orbital angular momentum $L_{in}$. Green colors represent binaries whose $S_2$ alignment with respect to $L_{in}$ can not be established or differentiated. Different markers are used to denote alignment type between spins $S_1$ and $S_2$. }    
        \label{fig7:a}
	\end{subfigure}
	\begin{subfigure}{.49\textwidth}
		\centering
		\includegraphics[width=\linewidth]{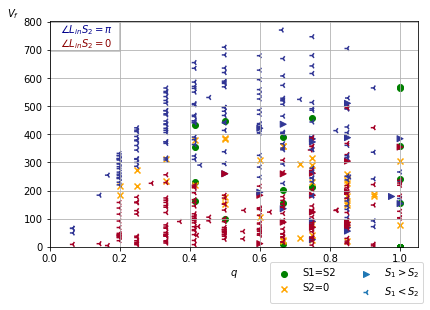}
		\caption{Red and blue colors denote alignment of spin $S_2$ with respect to orbital angular momentum $L_{in}$. Yellow color has $S_2=0$ and thus no alignment can be establish with respect to $L_{in}$. Green dots are indifferent as $S_1=S_2$. Markers clear out the bigger value of the spin in simulations.}
        \label{fig7:b}
   \end{subfigure}	
	
	\caption{Correlation between final velocity $V_f$ and mass ratio $q$ for aligned binaries.}
	\label{fig7}
\end{figure*}

In Fig.\ref{fig7} we outline the dependence of the final center of mass velocity $V_f$ with the initial mass ratio $q$. Figures (a) and (b) highlight different initial configurations for magnitudes and spins directions. In figure (c) a colorbar is used to show the highest luminosity released for each configuration. Over all the 407 aligned simulations, the norm of $S_2$ sweeps values in [0,0.7] interval and the absolute value of $S_1$ in [0.,0.25] interval. 

Fig.\ref{fig7:a} shows the highest final velocity $V_f= 771.78\ km/s$ for the quotient $q=0.6628$ in the whole class of aligned-spins binaries. In the same plot we find that top velocities can only be achieved for BBHs with anti-aligned spins ($\angle S_1 S_2=\pi$). This property has already been reported several times \cite{pollney2007recoil,koppitz2007recoil}. On the other hand, there seems to be no clear picture of the influence between the spin alignment of $S_1$ and $S_2$ in lower final velocities. 

Furthermore, it can be seen in Fig.\ref{fig7:b} that when the initial orbital angular momentum $L_{in}$ is anti-aligned with the initial spin $S_2$ ($\angle L_{in} S_2=\pi$), higher final velocities are reached. In short, it seems that the higher absolute value spins, in this case $S_2$, should be anti-aligned with initial orbital angular momentum $L_{in}$ and the initial lower spin, in this case $S_1$, to achieve the maximum velocities for the aligned group.

\begin{figure}[htb]
	\centering
	\includegraphics[scale=0.5]{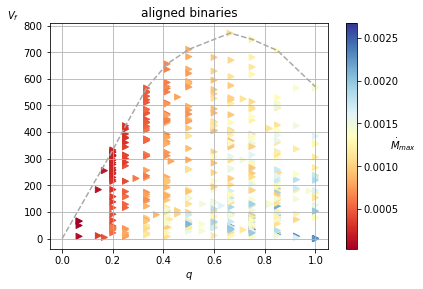}
	\caption{Correlation between $V_f$ and $q$ for aligned binaries. Colorbar shows levels of peak energy loss $\dot{M}_{max}$. Dashed line indicates border limit to final velocities (in $km/s$).}
    \label{fig7:c}
\end{figure}

The correlation between final velocities $V_f$ and maximum rate of mass loss $\dot{M}_{max}$ is shown in Fig.\ref{fig7:c}. A similar pattern discussed for NS binaries \ref{Non-spinning binaries} is also present. The maximum possible velocity versus q has a local (and absolute) maximum at $q\sim 0.66 $. Final velocities vanishes at $q=0$, reach a peak value and then decrease at $q=1$. Moreover, energy loss also vanish at $q=0$ which indicates that both the NS and A classes are very similar at this part of the mass ratio range.

At the other side of the range, the equal mass pattern $q=1$ is quite interesting. Like the NS case, the maximum energy loss and vanishing final velocity also occurs at $q=1$. However, unlike the NS case, there are now non-vanishing final velocities in the EM-A configuration, a big difference from the NS case.

\begin{figure}[htb]
	\centering
	\includegraphics[scale=0.5]{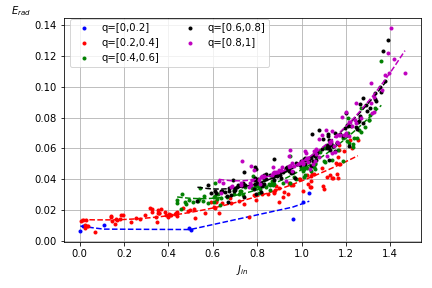}
	\caption{Correlation between initial total angular momentum $J_{in}$ and total radiation emitted $E_{rad}$ for aligned binaries. Mass ratio intervals have been divided and selected to be studied and are illustrated with different colors}
	\label{fig8}
\end{figure}

The relation of the energy loss is then analyzed with respect to the initial total angular momentum $J_{in}$. In Fig. \ref{fig8}, the dependence between the initial total angular momentum $J_{in}$ and total radiated energy $E_{rad}$ is shown for five mass range ratios: $0<q\le0.2$, $0.2<q\le0.4$, $0.4<q\le0.6$, $0.6<q\le0.8$ and $0.8<q\le1$. We fit a quadratic polynomial model for each range which could be useful for estimating energy on astrophysical scenarios or also for future comparisons. We provided the coefficients of the fit for each range of masses in Table \ref{(A)Erad_vs_Jin_coef}. 

\begin{table}[ht]
\centering
\begin{tabular}{|c|c|c|c|}
\hline
$q$   &   $a_0$ &        $a_1$ &       $a_2$  \\
\hline
$0<q\le0.2$ & $0.037 \pm 0.016$ & $-0.023 \pm 0.017$ & $0.0095 \pm 0.0031$\\   
$0.2<q\le0.4$ & $0.032 \pm 0.004$ & $-0.007 \pm 0.005$ &  $0.0138 \pm 0.0013$ \\
$0.4<q\le0.6$ & $0.093 \pm 0.009$ & $-0.102 \pm 0.015$ &  $0.0553\pm 0.0065$ \\
$0.6<q\le0.8$ & $0.12\pm 0.01$ & $-0.156\pm 0.019$ &  $0.082 \pm 0.009$ \\
$0.8<q\le1$ & $0.149\pm0.014$ & $-0.212\pm 0.029$ &  $0.114\pm0.014$ \\ 
\hline
\end{tabular}
\caption{Coefficients from fitted second-degree polynomials in Fig.\ref{fig8}}
\label{(A)Erad_vs_Jin_coef}
\end{table}

To end the section of aligned-spins BBHs and for comparison purposes, we will restrict our analysis to $q=1$ aligned BBHs and employ Reisswig \textit{et al.} model from Ref. \cite{reisswig2009gravitational}. The authors have shown that radiated energy via gravitational waves from infinity of equal-mass binaries with aligned spins can be estimated by a quadratic polynomial on the average initial spin $\bar{\chi}= (\chi_1+\chi_2)/2$, where $\chi_1$ and $\chi_2$ are the projections of the initial spin in the $L_{in}$ direction. Therefore, the model reads
\begin{equation}
    E_{rad}=a_0+a_1 \bar{\chi} +a_2 \bar{\chi}^2.
\end{equation}

Our fit to the data give us the following vector coefficients
\begin{equation} \label{coefNuestro}
    \vec{a}=\begin{pmatrix}
    a_0\\
    a_1\\
    a_2
    \end{pmatrix}=
    \begin{pmatrix}
    0.051\pm 0.001\\
    0.040\pm 0.002\\
    0.029 \pm 0.003
    \end{pmatrix} ,
\end{equation}
whereas the corresponding coefficients in Ref. \cite{reisswig2009gravitational}  yield
\begin{equation} \label{coefReisswig}
    \vec{p}=\begin{pmatrix}
    p_0\\
    p_1\\
    p_2
    \end{pmatrix}=
    \begin{pmatrix}
    0.036 \pm 0.003 \\
    0.030\pm 0.006\\
    0.02\pm 0.01
    \end{pmatrix}
\end{equation}.

\begin{figure}[htb]
	\centering
	\includegraphics[scale=0.5]{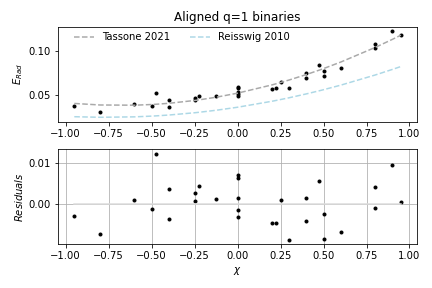}
	\caption{Correlation between initial total spin $\bar{\chi}$ and total radiation emitted $E_{rad}$. Dashed grey line shows the fit to our data and below its residuals. Dashed light blue line shows Reisswig polynomial for comparison purposes.}
	\label{fig8.5}
\end{figure}

 We see in Fig.\ref{fig8.5} there is a notable difference between the polynomials. The difference could be attributed to the fact that the parameters for the used simulations differ significantly and thus the final the $E_{rad}$. Particularly, no comment is said about initial orbital angular momentum $L_{in}$ in Ref. \cite{reisswig2009gravitational} which can generate different outcomes in the radiated energy for the same values of the initial spins $\chi_1$,$\chi_2$. However, it could be also that radiated energy predicted for aligned BBHs in the framework of KQ formalism is slightly higher than radiated energy obtained using the local quantities such as the Christodoulou mass in numerical relativity. 
 
 The largest error is in 2nd-order coefficient and is $\sim 10\%$. We find the maximum radiated energy evaluating at $a=1$ in the polynomial and we find the value $E_{rad}(1)=12.2\%$. The limit is closer to $E_{rad}(1)=11.3\%$ reported in \cite{hemberger2013} than $E_{rad}(1)=9.9\%$ reported in \cite{reisswig2009gravitational}. The maximum radiated energy is also below the value $E_{rad} \sim 14\% $ found in Ref. \cite{sperhake2008high} for head-on collision of two highly boosted EM-NS black holes.
 
\subsection{Precessing spins binaries}
We provide in this section a deeper analysis on the parameters for the precessing spins group. For this group of BBH, initial spins $S_1$ and $S_2$ will, in general, not be aligned nor even along $\textbf{L}_{in}$ axis. Different behaviour with respect to the previous groups mentioned in Secs.\ref{Non-spinning binaries} and \ref{Aligned binaries} is expected as non aligned configurations are known to present chaotic behaviour \cite{kozameh2020spin}.

\begin{figure*}[htb]
	\centering
	\begin{subfigure}{.49\textwidth}
		\centering
		\includegraphics[width=\linewidth]{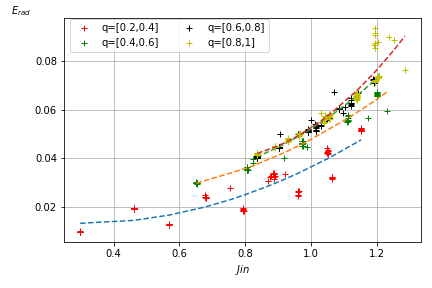}
		\caption{Correlation between initial total angular momentum $J_{in}$ and total radiation emitted $E_{rad}$.}
        \label{fig9:a}
	\end{subfigure}
	\begin{subfigure}{.49\textwidth}
		\centering
		\includegraphics[width=\linewidth]{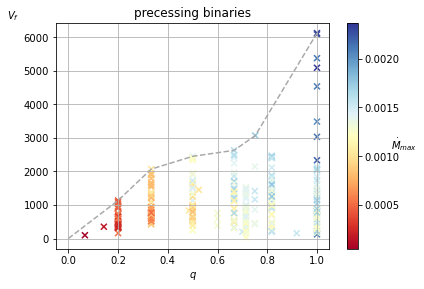}
        \caption{Correlation between final velocity $V_f$ (in $km/s$) with the initial mass ratio $q$. Colorbar shows different levels of peak energy loss.}
        \label{fig9:b}
	\end{subfigure}
	\caption{Detailed description of the many variable dependence for the precessing binaries.}
	\label{fig9}
\end{figure*}

We first study, as in previous sections, the dependence of the final velocity $V_f$ with the initial mass ratio $q$. In Fig.\ref{fig9:b} one can appreciate that the most notable difference in the behaviour of $V_f$ with respect to the non-spinning or aligned BBHs is the achievement of highest velocities for equal mass binaries. By contrast, $\dot{M}_{max}$ behaves equally as in other classes, i.e., systems with $q=0$ have no energy emission as expected and $q=1$ are the most energetic BBHs. The rate of energy loss increment as the mass ratio does.

On the left side, Fig.\ref{fig9:a} shows $E_{rad}$ vs initial total angular momentum $J_{in}$. We used the same model as with aligned and non-spinning classes derived in Appendix \ref{AppendixA} to fit the data, i.e., a 2nd degree polynomials for each range of masses.  The coefficients of the polynomials are listed in table \ref{(P)Erad_vs_Jin_coef}. Even though second degree polynomials may not be the best fit for the data, we prioritize its simplicity for describing the ascending behaviour of each mass range in the precessing group. Note also that range $0<q<0.2$ is missing in plot \ref{fig9:a} due to the lack of enough data to provide a representative curve in that interval.

\begin{table}[htb]
\begin{tabular}{|c|c|c|c|}
\hline
$q$   &   $a_2$ &        $a_1$ &       $a_0$  \\
\hline
$0.2<q\le0.4$ & $0.040\pm 0.009$ & $-0.020\pm 0.014$ & $0.017\pm 0.005$\\
$0.4<q\le0.6$ & $0.070 \pm 0.013$ & $-0.071 \pm 0.025$ & $0.046 \pm 0.011$ \\
$0.6<q\le0.8$ & $0.089\pm0.024$ & $-0.097\pm0.048$ &  $0.059\pm0.024$ \\
$0.8<q\le1$ & $0.180\pm0.038$ & $-0.271\pm0.081$ &  $0.143\pm0.041$\\
\hline
\end{tabular}
\caption{Coefficients of second-degree polynomials.}
\label{(P)Erad_vs_Jin_coef} 
\end{table}

\section{\label{Discussion}Discussion and Conclusion}

In this work, we have studied final velocities of the remnant black hole for the non-spinning (NS), aligned (A) and precessing (P) groups. We have shown that final velocities find their maximum value for different mass ratios in  each group: $V_f\approx 310 \ km/s$ for $q=0.4$ in NS, $V_f = 771\ km/s$ for $q=0.66$ in A, and $V_f = 6120.15\ km/s$ for $q=1$ in P. We have given the final velocities in Bondi time. For comparison purposes, we should divide our final velocities by a $\sqrt{2}$ factor, as explained in Sec.\ref{General results}. Thus, the maximum final velocity in the precessing group is $V_{f}=4327.59\ km/s$, which is slightly lower than the final kick reported for this simulation in Rochester metadata ($v_{kick}= 4625.73\ km/s$).

We have found $V_f=219.20\ km/s$ at $q=0.4$ to be the maximum final velocity for NS group. This value is consistent with the reported maximum value $V_f=175\ km/s$  for the ratio $q\approx 0.36$ found in Ref. \cite{gonzalez2007}. For the NS group, we employed the Fitchett's recoil model to fit the set of NS simulations in the catalog. We obtained a fitting constant $a=16317.78\ km/s$ and associated error of $\sim 1.5\%$. The model can be used for posterior simulations comparison or to model astrophysical BBHs with negligible rotation. Likewise, we derived a formula to fit the correlation between $\dot{E}_{max}$ and $q$ using a phenomenological dimensionless parameter $b$ which should be greater than $2$. The value obtained $b=2.1867$ is a reasonable result.

For the A group, we found that the top final velocity is $V_f=545.73\ km/s$, which is consistent with the maximum value in the metadata $v_{kick}=500\ km/s$ and of the same order found in Ref.\cite{pollney2007recoil}, $V_f=448\ km/s$. Moreover, our results also show that final velocities are maximal when spins are antialigned. This result have been also reported in other works \cite{pollney2007recoil,koppitz2007recoil}.

We have also found that the highest attainable final speeds come from the EM-P group. According to the Rochester repository, if the remnant black hole is found with speeds higher than $800\ km/s$, then the BBH had precessing spins. Moreover, if the speed of the remnant BH is higher than $3000\ km/s$ then the inital binaries belong to the EM-P subclass.

We have also studied in depth the radiation for each class predicted by \cref{Erad}. For the NS group, we find that using the model derived in Appendix \ref{AppendixA},  a quadratic polynomial fits notably well the data with a coefficient $b=0.04976$ and error  $\sim 1.74\%$. The simplicity of the non-spinning binaries allows one to model the dependence of final variables with initial parameters and to obtain formulas with smaller errors than in other classes of BBHs.

In the group of aligned spins binaries we made a fit for different ranges of mass ratios and found higher error values than non-spinning binaries fit. The higher errors could be attributed to the coefficients dependence of the spin. Perhaps more complex models should be considered to reduce the coefficient errors. Despite that, the coefficients calculated still provide a simple and useful model to the relation between $E_{rad}$ and $ J_{in}$.
For the case of $q=1$ aligned binaries we fit again the radiated energy but use the variable $\bar{\chi}$ instead of $J_{in}$. The plot in terms of the variable $\bar{\chi}$ allows us to compare the predicted radiated energy with other works. We see the fit is consistent although barely higher than most reported radiated energies. This could be a consequence of defining \ref{Erad} with Bondi mass. A more thorough analysis on the difference between radiated energy defined via Eq.(\ref{Mdot}) and via Christodoulou mass should be made, and we left its study for future work.

We found similar results for the precessing group. There are some ranges of mass ratio where the fits have higher error in the linear coefficients (up to $\sim 70 \%$) and data is not enough for fitting low mass ratio binaries, i.e., for $0<q<0.2$.

We now make some general remarks related to this work:
\begin{itemize}
    \item Reisswig \textit{et al} \cite{reisswig2009gravitational} have shown that the gravitational wave pattern for EM-NS and EM-A are quite similar. Since the numerical evolution for EM-NS binaries are much simpler than the EM-A subclass, they argue that one could simply use the NS class to obtain the waveforms for both subgroups.
    On the other hand, our results show that the final velocities for the EM-NS and EM-A sub classes are completely different. Whereas the former class have vanishing recoil velocities, EM-A cases do not. This is a distinctive difference between the two subclasses and helps to distinguish between them. Thus, even though the waveform templates are virtually the same for these sub classes, the numerical evolution of the EM-A subclass yields valuable information of the binary coalescence.
   \item Figs.\ref{fig8} and \ref{fig9:a} represent truly global results. The plot of $E_{rad}$ vs $J_{in}$ is motivated by the following. Both variables can be obtained at null infinity without knowledge of the BH masses, spins and orbital angular momentum. To obtain $J_{in}$ we only need the knowledge of the center of mass position and the definition of total angular momentum, whereas $E_{rad}$ is computed with the use of the Bondi mass loss equation. This plot shows a clear correlation between $E_{rad}$ and $J_{in}$. Using a quadratic fit we obtained an empirical relationship for the coefficients. The results in this works show that the initial total angular momentum $J_{in}$ is a relevant variable in the parameter space to analyze the different numerical evolutions. It is fair to ask why there should be a quadratic dependence on the initial total intrinsic angular momentum of the BBH system. The answer lies in the formula for $E_{rad}$, and it depends on quadratic gravitational radiation terms. At the same time, the radiation depends on the total initial angular momentum and it is conserved if we only keep up to quadratic terms in the formula for $E_{rad}$ since $\dot{J}$ is quadratic in the shear. Thus, we should expect this dependency in the BBH coalescence. Note the same results for $E_{rad}$ applies also for $\dot{M}_{max}$ as a linear relation between them was shown in \cref{fig3}. 
   
   \item The angular momentum loss $\Delta S$ only depends on the amount of angular momentum radiated away, and it is computed with knowledge of the available radiation data at null infinity. The angular momentum loss is still a main reason of concern in our formalism. As previously shown in \cite{tassone2021}, there is a discrepancy of the predicted intrinsic angular momentum with respect to the predicted values in RIT catalog, which is a subject to be studied in future works. Of course, a physically relevant definition of intrinsic angular momentum is a difficult task but the one we provide appears to be free from ambiguities. Despite that, we have shown throughout this paper that the equations of motion for energy, linear momentum, center of mass position and final velocity [\cref{Mdot,Pdot,Ddot,D}] employed in the framework of KQ formalism are consistent with the general literature. 
\end{itemize}

\begin{acknowledgments}
This research has been supported by grants from CONICET and the Agencia Nacional de Promoci\'on de la Investigaci\'on, el Desarrollo Tecnol\'ogico y la Innovaci\'on.
\end{acknowledgments}

\bibliography{apssamp}

\providecommand{\noopsort}[1]{}\providecommand{\singleletter}[1]{#1}%
\begin{thebibliography}{34}%
\makeatletter
\providecommand \@ifxundefined [1]{%
 \@ifx{#1\undefined}
}%
\providecommand \@ifnum [1]{%
 \ifnum #1\expandafter \@firstoftwo
 \else \expandafter \@secondoftwo
 \fi
}%
\providecommand \@ifx [1]{%
 \ifx #1\expandafter \@firstoftwo
 \else \expandafter \@secondoftwo
 \fi
}%
\providecommand \natexlab [1]{#1}%
\providecommand \enquote  [1]{``#1''}%
\providecommand \bibnamefont  [1]{#1}%
\providecommand \bibfnamefont [1]{#1}%
\providecommand \citenamefont [1]{#1}%
\providecommand \href@noop [0]{\@secondoftwo}%
\providecommand \href [0]{\begingroup \@sanitize@url \@href}%
\providecommand \@href[1]{\@@startlink{#1}\@@href}%
\providecommand \@@href[1]{\endgroup#1\@@endlink}%
\providecommand \@sanitize@url [0]{\catcode `\\12\catcode `\$12\catcode
  `\&12\catcode `\#12\catcode `\^12\catcode `\_12\catcode `\%12\relax}%
\providecommand \@@startlink[1]{}%
\providecommand \@@endlink[0]{}%
\providecommand \url  [0]{\begingroup\@sanitize@url \@url }%
\providecommand \@url [1]{\endgroup\@href {#1}{\urlprefix }}%
\providecommand \urlprefix  [0]{URL }%
\providecommand \Eprint [0]{\href }%
\providecommand \doibase [0]{https://doi.org/}%
\providecommand \selectlanguage [0]{\@gobble}%
\providecommand \bibinfo  [0]{\@secondoftwo}%
\providecommand \bibfield  [0]{\@secondoftwo}%
\providecommand \translation [1]{[#1]}%
\providecommand \BibitemOpen [0]{}%
\providecommand \bibitemStop [0]{}%
\providecommand \bibitemNoStop [0]{.\EOS\space}%
\providecommand \EOS [0]{\spacefactor3000\relax}%
\providecommand \BibitemShut  [1]{\csname bibitem#1\endcsname}%
\let\auto@bib@innerbib\@empty
\bibitem [{\citenamefont {Healy}\ and\ \citenamefont
  {Lousto}(2020)}]{healy2020}%
  \BibitemOpen
  \bibfield  {author} {\bibinfo {author} {\bibfnamefont {J.}~\bibnamefont
  {Healy}}\ and\ \bibinfo {author} {\bibfnamefont {C.~O.}\ \bibnamefont
  {Lousto}},\ }\bibfield  {title} {\bibinfo {title} {Third rit binary black
  hole simulations catalog},\ }\href@noop {} {\bibfield  {journal} {\bibinfo
  {journal} {Physical Review D}\ }\textbf {\bibinfo {volume} {102}},\ \bibinfo
  {pages} {104018} (\bibinfo {year} {2020})}\BibitemShut {NoStop}%
\bibitem [{\citenamefont {Hemberger}\ \emph {et~al.}(2013)\citenamefont
  {Hemberger}, \citenamefont {Lovelace}, \citenamefont {Loredo}, \citenamefont
  {Kidder}, \citenamefont {Scheel}, \citenamefont {Szil{\'a}gyi}, \citenamefont
  {Taylor},\ and\ \citenamefont {Teukolsky}}]{hemberger2013}%
  \BibitemOpen
  \bibfield  {author} {\bibinfo {author} {\bibfnamefont {D.~A.}\ \bibnamefont
  {Hemberger}}, \bibinfo {author} {\bibfnamefont {G.}~\bibnamefont {Lovelace}},
  \bibinfo {author} {\bibfnamefont {T.~J.}\ \bibnamefont {Loredo}}, \bibinfo
  {author} {\bibfnamefont {L.~E.}\ \bibnamefont {Kidder}}, \bibinfo {author}
  {\bibfnamefont {M.~A.}\ \bibnamefont {Scheel}}, \bibinfo {author}
  {\bibfnamefont {B.}~\bibnamefont {Szil{\'a}gyi}}, \bibinfo {author}
  {\bibfnamefont {N.~W.}\ \bibnamefont {Taylor}},\ and\ \bibinfo {author}
  {\bibfnamefont {S.~A.}\ \bibnamefont {Teukolsky}},\ }\bibfield  {title}
  {\bibinfo {title} {Final spin and radiated energy in numerical simulations of
  binary black holes with equal masses and equal, aligned or antialigned
  spins},\ }\href@noop {} {\bibfield  {journal} {\bibinfo  {journal} {Physical
  Review D}\ }\textbf {\bibinfo {volume} {88}},\ \bibinfo {pages} {064014}
  (\bibinfo {year} {2013})}\BibitemShut {NoStop}%
\bibitem [{\citenamefont {Gonz{\'a}lez}\ \emph {et~al.}(2007)\citenamefont
  {Gonz{\'a}lez}, \citenamefont {Hannam}, \citenamefont {Sperhake},
  \citenamefont {Bruegmann},\ and\ \citenamefont {Husa}}]{gonzalez2007}%
  \BibitemOpen
  \bibfield  {author} {\bibinfo {author} {\bibfnamefont {J.~A.}\ \bibnamefont
  {Gonz{\'a}lez}}, \bibinfo {author} {\bibfnamefont {M.}~\bibnamefont
  {Hannam}}, \bibinfo {author} {\bibfnamefont {U.}~\bibnamefont {Sperhake}},
  \bibinfo {author} {\bibfnamefont {B.}~\bibnamefont {Bruegmann}},\ and\
  \bibinfo {author} {\bibfnamefont {S.}~\bibnamefont {Husa}},\ }\bibfield
  {title} {\bibinfo {title} {Supermassive recoil velocities for binary
  black-hole mergers with antialigned spins},\ }\href@noop {} {\bibfield
  {journal} {\bibinfo  {journal} {Physical Review Letters}\ }\textbf {\bibinfo
  {volume} {98}},\ \bibinfo {pages} {231101} (\bibinfo {year}
  {2007})}\BibitemShut {NoStop}%
\bibitem [{\citenamefont {Campanelli}\ \emph
  {et~al.}(2007{\natexlab{a}})\citenamefont {Campanelli}, \citenamefont
  {Lousto}, \citenamefont {Zlochower},\ and\ \citenamefont
  {Merritt}}]{campanelli2007}%
  \BibitemOpen
  \bibfield  {author} {\bibinfo {author} {\bibfnamefont {M.}~\bibnamefont
  {Campanelli}}, \bibinfo {author} {\bibfnamefont {C.}~\bibnamefont {Lousto}},
  \bibinfo {author} {\bibfnamefont {Y.}~\bibnamefont {Zlochower}},\ and\
  \bibinfo {author} {\bibfnamefont {D.}~\bibnamefont {Merritt}},\ }\bibfield
  {title} {\bibinfo {title} {Large merger recoils and spin flips from generic
  black hole binaries},\ }\href@noop {} {\bibfield  {journal} {\bibinfo
  {journal} {The Astrophysical Journal Letters}\ }\textbf {\bibinfo {volume}
  {659}},\ \bibinfo {pages} {L5} (\bibinfo {year}
  {2007}{\natexlab{a}})}\BibitemShut {NoStop}%
\bibitem [{\citenamefont {Campanelli}\ \emph
  {et~al.}(2007{\natexlab{b}})\citenamefont {Campanelli}, \citenamefont
  {Lousto}, \citenamefont {Zlochower},\ and\ \citenamefont
  {Merritt}}]{campanelli2007maximum}%
  \BibitemOpen
  \bibfield  {author} {\bibinfo {author} {\bibfnamefont {M.}~\bibnamefont
  {Campanelli}}, \bibinfo {author} {\bibfnamefont {C.~O.}\ \bibnamefont
  {Lousto}}, \bibinfo {author} {\bibfnamefont {Y.}~\bibnamefont {Zlochower}},\
  and\ \bibinfo {author} {\bibfnamefont {D.}~\bibnamefont {Merritt}},\
  }\bibfield  {title} {\bibinfo {title} {Maximum gravitational recoil},\
  }\href@noop {} {\bibfield  {journal} {\bibinfo  {journal} {Physical Review
  Letters}\ }\textbf {\bibinfo {volume} {98}},\ \bibinfo {pages} {231102}
  (\bibinfo {year} {2007}{\natexlab{b}})}\BibitemShut {NoStop}%
\bibitem [{\citenamefont {Healy}\ \emph {et~al.}(2014)\citenamefont {Healy},
  \citenamefont {Lousto},\ and\ \citenamefont {Zlochower}}]{healy2014}%
  \BibitemOpen
  \bibfield  {author} {\bibinfo {author} {\bibfnamefont {J.}~\bibnamefont
  {Healy}}, \bibinfo {author} {\bibfnamefont {C.~O.}\ \bibnamefont {Lousto}},\
  and\ \bibinfo {author} {\bibfnamefont {Y.}~\bibnamefont {Zlochower}},\
  }\bibfield  {title} {\bibinfo {title} {{Remnant mass, spin, and recoil from
  spin aligned black-hole binaries}},\ }\href
  {https://doi.org/10.1103/PhysRevD.90.104004} {\bibfield  {journal} {\bibinfo
  {journal} {Phys. Rev.}\ }\textbf {\bibinfo {volume} {D90}},\ \bibinfo {pages}
  {104004} (\bibinfo {year} {2014})},\ \Eprint
  {https://arxiv.org/abs/1406.7295} {arXiv:1406.7295 [gr-qc]} \BibitemShut
  {NoStop}%
\bibitem [{\citenamefont {Healy}\ and\ \citenamefont
  {Lousto}(2018)}]{healy2018hangup}%
  \BibitemOpen
  \bibfield  {author} {\bibinfo {author} {\bibfnamefont {J.}~\bibnamefont
  {Healy}}\ and\ \bibinfo {author} {\bibfnamefont {C.~O.}\ \bibnamefont
  {Lousto}},\ }\bibfield  {title} {\bibinfo {title} {Hangup effect in unequal
  mass binary black hole mergers and further studies of their gravitational
  radiation and remnant properties},\ }\href@noop {} {\bibfield  {journal}
  {\bibinfo  {journal} {Physical Review D}\ }\textbf {\bibinfo {volume} {97}},\
  \bibinfo {pages} {084002} (\bibinfo {year} {2018})}\BibitemShut {NoStop}%
\bibitem [{\citenamefont {Campanelli}\ \emph {et~al.}(2006)\citenamefont
  {Campanelli}, \citenamefont {Lousto},\ and\ \citenamefont
  {Zlochower}}]{campanelli2006spinning}%
  \BibitemOpen
  \bibfield  {author} {\bibinfo {author} {\bibfnamefont {M.}~\bibnamefont
  {Campanelli}}, \bibinfo {author} {\bibfnamefont {C.~O.}\ \bibnamefont
  {Lousto}},\ and\ \bibinfo {author} {\bibfnamefont {Y.}~\bibnamefont
  {Zlochower}},\ }\bibfield  {title} {\bibinfo {title} {Spinning-black-hole
  binaries: The orbital hang-up},\ }\href@noop {} {\bibfield  {journal}
  {\bibinfo  {journal} {Physical Review D}\ }\textbf {\bibinfo {volume} {74}},\
  \bibinfo {pages} {041501} (\bibinfo {year} {2006})}\BibitemShut {NoStop}%
\bibitem [{\citenamefont {Zlochower}\ and\ \citenamefont
  {Lousto}(2015)}]{zlochower2015}%
  \BibitemOpen
  \bibfield  {author} {\bibinfo {author} {\bibfnamefont {Y.}~\bibnamefont
  {Zlochower}}\ and\ \bibinfo {author} {\bibfnamefont {C.~O.}\ \bibnamefont
  {Lousto}},\ }\bibfield  {title} {\bibinfo {title} {Modeling the remnant mass,
  spin, and recoil from unequal-mass, precessing black-hole binaries: The
  intermediate mass ratio regime},\ }\href
  {https://doi.org/10.1103/PhysRevD.92.024022} {\bibfield  {journal} {\bibinfo
  {journal} {Phys. Rev. D}\ }\textbf {\bibinfo {volume} {92}},\ \bibinfo
  {pages} {024022} (\bibinfo {year} {2015})}\BibitemShut {NoStop}%
\bibitem [{\citenamefont {Rezzolla}\ \emph {et~al.}(2008)\citenamefont
  {Rezzolla}, \citenamefont {Barausse}, \citenamefont {Dorband}, \citenamefont
  {Pollney}, \citenamefont {Reisswig}, \citenamefont {Seiler},\ and\
  \citenamefont {Husa}}]{rezzolla2008final}%
  \BibitemOpen
  \bibfield  {author} {\bibinfo {author} {\bibfnamefont {L.}~\bibnamefont
  {Rezzolla}}, \bibinfo {author} {\bibfnamefont {E.}~\bibnamefont {Barausse}},
  \bibinfo {author} {\bibfnamefont {E.~N.}\ \bibnamefont {Dorband}}, \bibinfo
  {author} {\bibfnamefont {D.}~\bibnamefont {Pollney}}, \bibinfo {author}
  {\bibfnamefont {C.}~\bibnamefont {Reisswig}}, \bibinfo {author}
  {\bibfnamefont {J.}~\bibnamefont {Seiler}},\ and\ \bibinfo {author}
  {\bibfnamefont {S.}~\bibnamefont {Husa}},\ }\bibfield  {title} {\bibinfo
  {title} {Final spin from the coalescence of two black holes},\ }\href@noop {}
  {\bibfield  {journal} {\bibinfo  {journal} {Physical Review D}\ }\textbf
  {\bibinfo {volume} {78}},\ \bibinfo {pages} {044002} (\bibinfo {year}
  {2008})}\BibitemShut {NoStop}%
\bibitem [{\citenamefont {Hofmann}\ \emph {et~al.}(2016)\citenamefont
  {Hofmann}, \citenamefont {Barausse},\ and\ \citenamefont
  {Rezzolla}}]{hofmann2016final}%
  \BibitemOpen
  \bibfield  {author} {\bibinfo {author} {\bibfnamefont {F.}~\bibnamefont
  {Hofmann}}, \bibinfo {author} {\bibfnamefont {E.}~\bibnamefont {Barausse}},\
  and\ \bibinfo {author} {\bibfnamefont {L.}~\bibnamefont {Rezzolla}},\
  }\bibfield  {title} {\bibinfo {title} {The final spin from binary black holes
  in quasi-circular orbits},\ }\href@noop {} {\bibfield  {journal} {\bibinfo
  {journal} {The Astrophysical Journal Letters}\ }\textbf {\bibinfo {volume}
  {825}},\ \bibinfo {pages} {L19} (\bibinfo {year} {2016})}\BibitemShut
  {NoStop}%
\bibitem [{\citenamefont {Lousto}\ and\ \citenamefont
  {Zlochower}(2014)}]{lousto2014black}%
  \BibitemOpen
  \bibfield  {author} {\bibinfo {author} {\bibfnamefont {C.~O.}\ \bibnamefont
  {Lousto}}\ and\ \bibinfo {author} {\bibfnamefont {Y.}~\bibnamefont
  {Zlochower}},\ }\bibfield  {title} {\bibinfo {title} {Black hole binary
  remnant mass and spin: A new phenomenological formula},\ }\href@noop {}
  {\bibfield  {journal} {\bibinfo  {journal} {Physical Review D}\ }\textbf
  {\bibinfo {volume} {89}},\ \bibinfo {pages} {104052} (\bibinfo {year}
  {2014})}\BibitemShut {NoStop}%
\bibitem [{\citenamefont {Lousto}\ \emph {et~al.}(2010)\citenamefont {Lousto},
  \citenamefont {Campanelli}, \citenamefont {Zlochower},\ and\ \citenamefont
  {Nakano}}]{lousto2010remnant}%
  \BibitemOpen
  \bibfield  {author} {\bibinfo {author} {\bibfnamefont {C.~O.}\ \bibnamefont
  {Lousto}}, \bibinfo {author} {\bibfnamefont {M.}~\bibnamefont {Campanelli}},
  \bibinfo {author} {\bibfnamefont {Y.}~\bibnamefont {Zlochower}},\ and\
  \bibinfo {author} {\bibfnamefont {H.}~\bibnamefont {Nakano}},\ }\bibfield
  {title} {\bibinfo {title} {Remnant masses, spins and recoils from the merger
  of generic black hole binaries},\ }\href@noop {} {\bibfield  {journal}
  {\bibinfo  {journal} {Classical and Quantum Gravity}\ }\textbf {\bibinfo
  {volume} {27}},\ \bibinfo {pages} {114006} (\bibinfo {year}
  {2010})}\BibitemShut {NoStop}%
\bibitem [{\citenamefont {Blanchet}(2014)}]{blanchet2014gravitational}%
  \BibitemOpen
  \bibfield  {author} {\bibinfo {author} {\bibfnamefont {L.}~\bibnamefont
  {Blanchet}},\ }\bibfield  {title} {\bibinfo {title} {Gravitational radiation
  from post-newtonian sources and inspiralling compact binaries},\ }\href@noop
  {} {\bibfield  {journal} {\bibinfo  {journal} {Living reviews in relativity}\
  }\textbf {\bibinfo {volume} {17}},\ \bibinfo {pages} {1} (\bibinfo {year}
  {2014})}\BibitemShut {NoStop}%
\bibitem [{\citenamefont {Kozameh}\ and\ \citenamefont
  {Quiroga}(2016)}]{kozameh2016}%
  \BibitemOpen
  \bibfield  {author} {\bibinfo {author} {\bibfnamefont {C.~N.}\ \bibnamefont
  {Kozameh}}\ and\ \bibinfo {author} {\bibfnamefont {G.~D.}\ \bibnamefont
  {Quiroga}},\ }\bibfield  {title} {\bibinfo {title} {Center of mass and spin
  for isolated sources of gravitational radiation},\ }\href@noop {} {\bibfield
  {journal} {\bibinfo  {journal} {Physical Review D}\ }\textbf {\bibinfo
  {volume} {93}},\ \bibinfo {pages} {064050} (\bibinfo {year}
  {2016})}\BibitemShut {NoStop}%
\bibitem [{\citenamefont {Winicour}(1980)}]{held1980:3}%
  \BibitemOpen
  \bibfield  {author} {\bibinfo {author} {\bibfnamefont {J.}~\bibnamefont
  {Winicour}},\ }\bibfield  {title} {\bibinfo {title} {Angular momentum in
  general relativity},\ }in\ \href@noop {} {\emph {\bibinfo {booktitle}
  {General relativity and gravitation}}},\ Vol.~\bibinfo {volume} {2},\
  \bibinfo {editor} {edited by\ \bibinfo {editor} {\bibfnamefont
  {A.}~\bibnamefont {Held}}}\ (\bibinfo  {publisher} {Plenum Press},\ \bibinfo
  {address} {New York},\ \bibinfo {year} {1980})\BibitemShut {NoStop}%
\bibitem [{\citenamefont {Adamo}\ \emph {et~al.}(2012)\citenamefont {Adamo},
  \citenamefont {Newman},\ and\ \citenamefont {Kozameh}}]{adamo2012null}%
  \BibitemOpen
  \bibfield  {author} {\bibinfo {author} {\bibfnamefont {T.~M.}\ \bibnamefont
  {Adamo}}, \bibinfo {author} {\bibfnamefont {E.~T.}\ \bibnamefont {Newman}},\
  and\ \bibinfo {author} {\bibfnamefont {C.}~\bibnamefont {Kozameh}},\
  }\bibfield  {title} {\bibinfo {title} {Null geodesic congruences,
  asymptotically-flat spacetimes and their physical interpretation},\
  }\href@noop {} {\bibfield  {journal} {\bibinfo  {journal} {Living reviews in
  relativity}\ }\textbf {\bibinfo {volume} {15}},\ \bibinfo {pages} {1}
  (\bibinfo {year} {2012})}\BibitemShut {NoStop}%
\bibitem [{\citenamefont {Kozameh}\ \emph {et~al.}(2018)\citenamefont
  {Kozameh}, \citenamefont {Nieva},\ and\ \citenamefont
  {Quiroga}}]{kozameh2018}%
  \BibitemOpen
  \bibfield  {author} {\bibinfo {author} {\bibfnamefont {C.~N.}\ \bibnamefont
  {Kozameh}}, \bibinfo {author} {\bibfnamefont {J.~I.}\ \bibnamefont {Nieva}},\
  and\ \bibinfo {author} {\bibfnamefont {G.~D.}\ \bibnamefont {Quiroga}},\
  }\bibfield  {title} {\bibinfo {title} {Spin and center of mass comparison
  between the post-newtonian approach and the asymptotic formulation},\
  }\href@noop {} {\bibfield  {journal} {\bibinfo  {journal} {Physical Review
  D}\ }\textbf {\bibinfo {volume} {98}},\ \bibinfo {pages} {064005} (\bibinfo
  {year} {2018})}\BibitemShut {NoStop}%
\bibitem [{\citenamefont {Newman}\ and\ \citenamefont
  {Unti}(1963)}]{newman1963class}%
  \BibitemOpen
  \bibfield  {author} {\bibinfo {author} {\bibfnamefont {E.~T.}\ \bibnamefont
  {Newman}}\ and\ \bibinfo {author} {\bibfnamefont {T.}~\bibnamefont {Unti}},\
  }\bibfield  {title} {\bibinfo {title} {A class of null flat-space coordinate
  systems},\ }\href@noop {} {\bibfield  {journal} {\bibinfo  {journal} {Journal
  of Mathematical Physics}\ }\textbf {\bibinfo {volume} {4}},\ \bibinfo {pages}
  {1467} (\bibinfo {year} {1963})}\BibitemShut {NoStop}%
\bibitem [{\citenamefont {Kozameh}\ \emph {et~al.}(2020)\citenamefont
  {Kozameh}, \citenamefont {Nieva},\ and\ \citenamefont
  {Quiroga}}]{kozameh2020}%
  \BibitemOpen
  \bibfield  {author} {\bibinfo {author} {\bibfnamefont {C.}~\bibnamefont
  {Kozameh}}, \bibinfo {author} {\bibfnamefont {J.}~\bibnamefont {Nieva}},\
  and\ \bibinfo {author} {\bibfnamefont {G.}~\bibnamefont {Quiroga}},\
  }\bibfield  {title} {\bibinfo {title} {Relativistic center of mass in general
  relativity},\ }\href@noop {} {\bibfield  {journal} {\bibinfo  {journal}
  {Physical Review D}\ }\textbf {\bibinfo {volume} {101}},\ \bibinfo {pages}
  {024028} (\bibinfo {year} {2020})}\BibitemShut {NoStop}%
\bibitem [{\citenamefont {Thorne}(1980)}]{thorne1980multipole}%
  \BibitemOpen
  \bibfield  {author} {\bibinfo {author} {\bibfnamefont {K.~S.}\ \bibnamefont
  {Thorne}},\ }\bibfield  {title} {\bibinfo {title} {Multipole expansions of
  gravitational radiation},\ }\href@noop {} {\bibfield  {journal} {\bibinfo
  {journal} {Reviews of Modern Physics}\ }\textbf {\bibinfo {volume} {52}},\
  \bibinfo {pages} {299} (\bibinfo {year} {1980})}\BibitemShut {NoStop}%
\bibitem [{\citenamefont {Newman}\ and\ \citenamefont
  {Silva-Ortigoza}(2005)}]{newman2005}%
  \BibitemOpen
  \bibfield  {author} {\bibinfo {author} {\bibfnamefont {E.~T.}\ \bibnamefont
  {Newman}}\ and\ \bibinfo {author} {\bibfnamefont {G.}~\bibnamefont
  {Silva-Ortigoza}},\ }\bibfield  {title} {\bibinfo {title} {Tensorial spin-s
  harmonics},\ }\href@noop {} {\bibfield  {journal} {\bibinfo  {journal}
  {Classical and Quantum Gravity}\ }\textbf {\bibinfo {volume} {23}},\ \bibinfo
  {pages} {497} (\bibinfo {year} {2005})}\BibitemShut {NoStop}%
\bibitem [{\citenamefont {Tassone}()}]{code}%
  \BibitemOpen
  \bibfield  {author} {\bibinfo {author} {\bibfnamefont {E.~A.}\ \bibnamefont
  {Tassone}},\ }\href@noop {} {\bibinfo {title} {pykqevolution}},\ \bibinfo
  {howpublished}
  {\url{https://github.com/Emmatassone/pyKQEvolution}}\BibitemShut {NoStop}%
\bibitem [{\citenamefont {Tassone}\ \emph {et~al.}(2021)\citenamefont
  {Tassone}, \citenamefont {Mandrilli}, \citenamefont {Kozameh}, \citenamefont
  {Quiroga},\ and\ \citenamefont {Nieva}}]{tassone2021}%
  \BibitemOpen
  \bibfield  {author} {\bibinfo {author} {\bibfnamefont {E.~A.}\ \bibnamefont
  {Tassone}}, \bibinfo {author} {\bibfnamefont {P.~A.}\ \bibnamefont
  {Mandrilli}}, \bibinfo {author} {\bibfnamefont {C.~N.}\ \bibnamefont
  {Kozameh}}, \bibinfo {author} {\bibfnamefont {G.~D.}\ \bibnamefont
  {Quiroga}},\ and\ \bibinfo {author} {\bibfnamefont {J.~I.}\ \bibnamefont
  {Nieva}},\ }\bibfield  {title} {\bibinfo {title} {Numerical evolution of the
  center of mass and angular momentum for binary black holes},\ }\href
  {https://doi.org/10.1103/PhysRevD.104.084038} {\bibfield  {journal} {\bibinfo
   {journal} {Phys. Rev. D}\ }\textbf {\bibinfo {volume} {104}},\ \bibinfo
  {pages} {084038} (\bibinfo {year} {2021})}\BibitemShut {NoStop}%
\bibitem [{\citenamefont {Reisswig}\ and\ \citenamefont
  {Pollney}(2011)}]{reisswig2011notes}%
  \BibitemOpen
  \bibfield  {author} {\bibinfo {author} {\bibfnamefont {C.}~\bibnamefont
  {Reisswig}}\ and\ \bibinfo {author} {\bibfnamefont {D.}~\bibnamefont
  {Pollney}},\ }\bibfield  {title} {\bibinfo {title} {Notes on the integration
  of numerical relativity waveforms},\ }\href@noop {} {\bibfield  {journal}
  {\bibinfo  {journal} {Classical and Quantum Gravity}\ }\textbf {\bibinfo
  {volume} {28}},\ \bibinfo {pages} {195015} (\bibinfo {year}
  {2011})}\BibitemShut {NoStop}%
\bibitem [{\citenamefont {numerical~realtivity group}()}]{RITcatalog}%
  \BibitemOpen
  \bibfield  {author} {\bibinfo {author} {\bibfnamefont {R.}~\bibnamefont
  {numerical~realtivity group}},\ }\href@noop {} {\bibinfo {title} {Catalog of
  numerical simulations}},\ \bibinfo {howpublished}
  {\url{https://ccrg.rit.edu/numerical-simulations}}\BibitemShut {NoStop}%
\bibitem [{\citenamefont {Mandrilli}\ \emph {et~al.}(2020)\citenamefont
  {Mandrilli}, \citenamefont {Nieva},\ and\ \citenamefont
  {Quiroga}}]{mandrilli2020}%
  \BibitemOpen
  \bibfield  {author} {\bibinfo {author} {\bibfnamefont {P.}~\bibnamefont
  {Mandrilli}}, \bibinfo {author} {\bibfnamefont {J.}~\bibnamefont {Nieva}},\
  and\ \bibinfo {author} {\bibfnamefont {G.}~\bibnamefont {Quiroga}},\
  }\bibfield  {title} {\bibinfo {title} {Correspondence between tensorial
  spin-s and spin-weighted spherical harmonics},\ }\href@noop {} {\bibfield
  {journal} {\bibinfo  {journal} {General Relativity and Gravitation}\ }\textbf
  {\bibinfo {volume} {52}},\ \bibinfo {pages} {1} (\bibinfo {year}
  {2020})}\BibitemShut {NoStop}%
\bibitem [{\citenamefont {Fitchett}(1983)}]{fitchett1983influence}%
  \BibitemOpen
  \bibfield  {author} {\bibinfo {author} {\bibfnamefont {M.}~\bibnamefont
  {Fitchett}},\ }\bibfield  {title} {\bibinfo {title} {The influence of
  gravitational wave momentum losses on the centre of mass motion of a
  newtonian binary system},\ }\href@noop {} {\bibfield  {journal} {\bibinfo
  {journal} {Monthly Notices of the Royal Astronomical Society}\ }\textbf
  {\bibinfo {volume} {203}},\ \bibinfo {pages} {1049} (\bibinfo {year}
  {1983})}\BibitemShut {NoStop}%
\bibitem [{\citenamefont {Kozameh}\ and\ \citenamefont
  {Quiroga}(2020)}]{kozameh2020spin}%
  \BibitemOpen
  \bibfield  {author} {\bibinfo {author} {\bibfnamefont {C.}~\bibnamefont
  {Kozameh}}\ and\ \bibinfo {author} {\bibfnamefont {G.}~\bibnamefont
  {Quiroga}},\ }\bibfield  {title} {\bibinfo {title} {Spin stability of a
  binary black hole coalescence},\ }\href@noop {} {\bibfield  {journal}
  {\bibinfo  {journal} {Physical Review D}\ }\textbf {\bibinfo {volume}
  {102}},\ \bibinfo {pages} {064015} (\bibinfo {year} {2020})}\BibitemShut
  {NoStop}%
\bibitem [{\citenamefont {Pollney}\ \emph {et~al.}(2007)\citenamefont
  {Pollney}, \citenamefont {Reisswig}, \citenamefont {Rezzolla}, \citenamefont
  {Szil{\'a}gyi}, \citenamefont {Ansorg}, \citenamefont {Deris}, \citenamefont
  {Diener}, \citenamefont {Dorband}, \citenamefont {Koppitz}, \citenamefont
  {Nagar} \emph {et~al.}}]{pollney2007recoil}%
  \BibitemOpen
  \bibfield  {author} {\bibinfo {author} {\bibfnamefont {D.}~\bibnamefont
  {Pollney}}, \bibinfo {author} {\bibfnamefont {C.}~\bibnamefont {Reisswig}},
  \bibinfo {author} {\bibfnamefont {L.}~\bibnamefont {Rezzolla}}, \bibinfo
  {author} {\bibfnamefont {B.}~\bibnamefont {Szil{\'a}gyi}}, \bibinfo {author}
  {\bibfnamefont {M.}~\bibnamefont {Ansorg}}, \bibinfo {author} {\bibfnamefont
  {B.}~\bibnamefont {Deris}}, \bibinfo {author} {\bibfnamefont
  {P.}~\bibnamefont {Diener}}, \bibinfo {author} {\bibfnamefont {E.~N.}\
  \bibnamefont {Dorband}}, \bibinfo {author} {\bibfnamefont {M.}~\bibnamefont
  {Koppitz}}, \bibinfo {author} {\bibfnamefont {A.}~\bibnamefont {Nagar}},
  \emph {et~al.},\ }\bibfield  {title} {\bibinfo {title} {Recoil velocities
  from equal-mass binary black-hole mergers: a systematic investigation of
  spin-orbit aligned configurations},\ }\href@noop {} {\bibfield  {journal}
  {\bibinfo  {journal} {Physical Review D}\ }\textbf {\bibinfo {volume} {76}},\
  \bibinfo {pages} {124002} (\bibinfo {year} {2007})}\BibitemShut {NoStop}%
\bibitem [{\citenamefont {Koppitz}\ \emph {et~al.}(2007)\citenamefont
  {Koppitz}, \citenamefont {Pollney}, \citenamefont {Reisswig}, \citenamefont
  {Rezzolla}, \citenamefont {Thornburg}, \citenamefont {Diener},\ and\
  \citenamefont {Schnetter}}]{koppitz2007recoil}%
  \BibitemOpen
  \bibfield  {author} {\bibinfo {author} {\bibfnamefont {M.}~\bibnamefont
  {Koppitz}}, \bibinfo {author} {\bibfnamefont {D.}~\bibnamefont {Pollney}},
  \bibinfo {author} {\bibfnamefont {C.}~\bibnamefont {Reisswig}}, \bibinfo
  {author} {\bibfnamefont {L.}~\bibnamefont {Rezzolla}}, \bibinfo {author}
  {\bibfnamefont {J.}~\bibnamefont {Thornburg}}, \bibinfo {author}
  {\bibfnamefont {P.}~\bibnamefont {Diener}},\ and\ \bibinfo {author}
  {\bibfnamefont {E.}~\bibnamefont {Schnetter}},\ }\bibfield  {title} {\bibinfo
  {title} {Recoil velocities from equal-mass binary-black-hole mergers},\
  }\href@noop {} {\bibfield  {journal} {\bibinfo  {journal} {Physical Review
  Letters}\ }\textbf {\bibinfo {volume} {99}},\ \bibinfo {pages} {041102}
  (\bibinfo {year} {2007})}\BibitemShut {NoStop}%
\bibitem [{\citenamefont {Reisswig}\ \emph {et~al.}(2009)\citenamefont
  {Reisswig}, \citenamefont {Husa}, \citenamefont {Rezzolla}, \citenamefont
  {Dorband}, \citenamefont {Pollney},\ and\ \citenamefont
  {Seiler}}]{reisswig2009gravitational}%
  \BibitemOpen
  \bibfield  {author} {\bibinfo {author} {\bibfnamefont {C.}~\bibnamefont
  {Reisswig}}, \bibinfo {author} {\bibfnamefont {S.}~\bibnamefont {Husa}},
  \bibinfo {author} {\bibfnamefont {L.}~\bibnamefont {Rezzolla}}, \bibinfo
  {author} {\bibfnamefont {E.~N.}\ \bibnamefont {Dorband}}, \bibinfo {author}
  {\bibfnamefont {D.}~\bibnamefont {Pollney}},\ and\ \bibinfo {author}
  {\bibfnamefont {J.}~\bibnamefont {Seiler}},\ }\bibfield  {title} {\bibinfo
  {title} {Gravitational-wave detectability of equal-mass black-hole binaries
  with aligned spins},\ }\href@noop {} {\bibfield  {journal} {\bibinfo
  {journal} {Physical Review D}\ }\textbf {\bibinfo {volume} {80}},\ \bibinfo
  {pages} {124026} (\bibinfo {year} {2009})}\BibitemShut {NoStop}%
\bibitem [{\citenamefont {Sperhake}\ \emph {et~al.}(2008)\citenamefont
  {Sperhake}, \citenamefont {Cardoso}, \citenamefont {Pretorius}, \citenamefont
  {Berti},\ and\ \citenamefont {Gonzalez}}]{sperhake2008high}%
  \BibitemOpen
  \bibfield  {author} {\bibinfo {author} {\bibfnamefont {U.}~\bibnamefont
  {Sperhake}}, \bibinfo {author} {\bibfnamefont {V.}~\bibnamefont {Cardoso}},
  \bibinfo {author} {\bibfnamefont {F.}~\bibnamefont {Pretorius}}, \bibinfo
  {author} {\bibfnamefont {E.}~\bibnamefont {Berti}},\ and\ \bibinfo {author}
  {\bibfnamefont {J.~A.}\ \bibnamefont {Gonzalez}},\ }\bibfield  {title}
  {\bibinfo {title} {High-energy collision of two black holes},\ }\href@noop {}
  {\bibfield  {journal} {\bibinfo  {journal} {Physical review letters}\
  }\textbf {\bibinfo {volume} {101}},\ \bibinfo {pages} {161101} (\bibinfo
  {year} {2008})}\BibitemShut {NoStop}%
\bibitem [{\citenamefont {Kidder}(1995)}]{kidder1995coalescing}%
  \BibitemOpen
  \bibfield  {author} {\bibinfo {author} {\bibfnamefont {L.~E.}\ \bibnamefont
  {Kidder}},\ }\bibfield  {title} {\bibinfo {title} {Coalescing binary systems
  of compact objects to (post) 5/2-newtonian order. v. spin effects},\
  }\href@noop {} {\bibfield  {journal} {\bibinfo  {journal} {Physical Review
  D}\ }\textbf {\bibinfo {volume} {52}},\ \bibinfo {pages} {821} (\bibinfo
  {year} {1995})}\BibitemShut {NoStop}%
\end{thebibliography}%

\clearpage

\appendix
\section{Relation between $\dot{E}$, $J_{in}$ and $q$} \label{AppendixA}
Starting from the definition of the mass quadrupole moments, \cite{kidder1995coalescing} or \cite{blanchet2014gravitational}, we have the following expressions:
\begin{align}
I^{ij}_N&=\eta m x^{<i}x^{j>}\\
J^{ij}_N&=-\frac{\delta m}{m}x^{<i}L^{j>}+\frac{3}{2}\eta m x^{<i}(m_2\chi_2 - m_1\chi_1) ^{j>},
\end{align}
with the mass parameters given as $m = m_1 + m_2$, $\delta m = m_1-m_2$,$L^i$ the orbital angular momentum, and $\eta =m_1m_2/m^2$ the symmetric mass ratio.
For simplicity we consider the non spinning, equal mass case. Thus, $\delta m = 0$, $m=1$, and $\eta=\frac{1}{4}$.With these mass parameters, the quadrupole radiative moments are
\begin{align}
I^{ij}_N&=\eta m\big[ x^ix^j-\frac{1}{3}\delta_{ij}x^2\big]\\
J^{ij}_N&=0.
\end{align}
Explicitly, the non-zero radiative moments in the \textit{N} approximation are given by,

\begin{eqnarray}
I_N^{zz}&=&-\frac{1}{12} m r^2\\
I_N^{xx}&=&\frac{1}{24} m r^2[1+3\cos(2\omega t)]\\
I_N^{yy}&=&\frac{1}{24} m r^2[1-3\cos(2\omega t)]\\
I_N^{xy}&=&I_N^{yx}=\frac{1}{8}m r^2[\sin(2\omega t)]
\end{eqnarray}

Since the orbital angular momentum $L= \eta m r^2\omega$ is conserved in the \textit{N} approximation we obtain 
\begin{eqnarray}
I_N^{(3)zz}&=&0\\
I_N^{(3)xx}&=& 4 L_{in}\omega^2[\sin(2\omega t)]\\
I_N^{(3)yy}&=&-4 L_{in}\omega^2[\sin(2\omega t)]\\
I_N^{(3)xy}&=&I_N^{(3)yx}=-4L_{in}\omega^2[\cos(2\omega t)],
\end{eqnarray}
where the symbol (3) denotes three time derivatives. Using Eq. (\ref{Mdot}) and taking an average over a period, we have
\begin{align}
	\dot{M}&=-\frac{1}{10\sqrt{2}}\dot{\sigma}_{R}^{ij}\dot{\sigma}_{R}^{ij}\\
	       &=-\frac{32}{10\sqrt{2}}L_{in}^2\omega^4.
\end{align}
Thus,

\begin{equation}
	\dot{M}_{max}=-\frac{32}{10\sqrt{2}}\omega_{max}^4L_{in}^2.
\end{equation}

If the masses are not equal, the resulting formula for the non spinning subclass is given by
\begin{equation}\label{Lum-J}
	\dot{M}_{max}=-\frac{16}{10\sqrt{2}}\omega_{max}^4L_{in}^2\left( 1 + \frac{1}{36}(\frac{1-q}{1+q})^2(M_{in}\omega_{max})^{\frac{2}{3}}\right).
\end{equation}
The highest allowed value for $\omega$ in the PN approximation is $0.05$. However, The resulting numerical factor of $4\times 10^{-5}$ in the above equation should not be compared with the one obtained from the numerical evolution since we are in the PN approximation and the binaries are not even close to the coalescing time. However, it is important to see the quadratic dependence of the maximum luminosity on the initial value of the total angular momentum of the system.
If we want to apply this model to the other spinning classes (A or P), we use $L^i=J^i-(S_1+S_2)^i$, which gives us a second order polynomial with non vanishing zeroth and first order coefficients. Although the coefficients will depend on the initial spins ($S_1,S_2$), the model will be still useful to give a representative value of the dependence. 

One can also write the maximum luminosity in terms of the ratio $q$ by writing the orbital angular momentum as
\begin{equation}
    L=\eta m r^2\omega = \frac{q}{(1+q)^2}m r^2\omega.
\end{equation}
Inserting this equation into (\ref{Lum-J}) and using the circular orbit formula $$\omega^2r^2= \frac{m}{r}$$ one obtains
\begin{equation}\label{Lum-q}
	\dot{M}=-\frac{16}{10\sqrt{2}}\frac{q^2}{(1+q)^4}\left(\frac{m}{r}\right)^5\left( 1 + \frac{1}{36}\frac{m}{r}\left(\frac{1-q}{1+q}\right)^2\right).
\end{equation}
The last stable circular orbit before coalescence is given by 
$$
r_{stable}= 6 m,
$$
Thus, the luminosity at that particular orbit is given by 
\begin{equation}\label{Lum}
	\dot{M}=-\frac{16}{10\sqrt{2}}\frac{q^2}{(1+q)^4}\left(\frac{1}{6}\right)^5\left( 1 + \frac{1}{216}\left(\frac{1-q}{1+q}\right)^2\right).
\end{equation}
We now assume that the final details of the coalescence do not change the parameter dependence given by eq. (\ref{Lum}). However, the huge amount of energy must be taken into account by giving two phenomenological constants, one for each contribution of the mass and magnetic quadrupole terms. We thus assume that
\begin{equation}
	\dot{M}=-\frac{A}{1000}\frac{q^2}{(1+q)^4}\left( 1 + \frac{B}{216}\left(\frac{1-q}{1+q}\right)^2\right),
\end{equation}
and the values of $A$ and $B$ are found by fitting the formula with the numerical data.

One can also make a phenomenological correlation between the final value of the velocity $V_f$ o momentum $P_f$ and $q$. Note that Eq. (\ref{Pdot}) depends on the product of the real and imaginary part of $\sigma$. Thus, an analogous derivation to the one performed for $\dot{M}$ yields
\begin{equation}\label{Vfinal}
	V_f=a\eta^2 \frac{\delta m}{m} = a\frac{q^2}{(1+q)^4}\frac{1-q}{1+q},
\end{equation}
where the parameter $a$ can be obtained from the numerical plot. This last result has been previously derived in Ref. \cite{fitchett1983influence}.

\end{document}